\input{aipcheck}

\documentclass[
    ,final            
  ]
  {aipproc}

\layoutstyle{6x9}

\newcommand{\be}{\begin{equation}}
\newcommand{\ee}{\end{equation}}

\newcommand{\ba}{\begin{eqnarray}}
\newcommand{\ea}{\end{eqnarray}}

\newcommand{\Bf}{{magnetic field\,}}

\newcommand{\Ef}{{electric  field\,}}
\newcommand{\Efs}{{electric fields\,}}

\newcommand{\apj}{ApJ}

\begin{document}

\title{IC model of pulsar high energy emission}


\author{Maxim Lyutikov}{
  address={Department of Physics, Purdue University, 
 525 Northwestern Avenue,
West Lafayette, IN
47907-2036, USA
\\
and
\\
INAF - Osservatorio Astrofisico di Arcetri, 
Largo Enrico Fermi 5, I - 50125 Firenze,  Italia }
}

\begin{abstract}
We discuss growing evidence that pulsar high energy is emission is generated via Inverse Compton  mechanism. We reproduce  the broadband  spectrum of  Crab pulsar, from UV to very high energy gamma-rays - nearly ten decades in energy, within the framework of  the cyclotron-self-Compton model. Emission is produced  by  two counter-streaming beams within the outer gaps, at distances above  $\sim$ 20 NS radii. The outward moving beam  produces UV-$X$-ray photons via Doppler-booster cyclotron emission, and GeV photons  by Compton scattering the cyclotron photons produced by the inward going beam. The scattering occurs in the deep Klein-Nishina regime, whereby the  IC component provides a direct measurement of particle distribution within the magnetosphere.   The required plasma multiplicity is high, $\sim 10^6-10^7$, but is consistent with the average particle flux injected into the pulsar wind nebula. 

\end{abstract}

\maketitle

\section{Evidence in favor of IC scattering as the main source of high energy photons}
The pulsar high energy emission is a complicated  unsolved problem in high energy astrophysics which has been  been under intensive study for nearly four decades.  {\it Geometrical} models, based on the idea of the outer gap \citep{1986ApJ...300..500C}, are very successful in explaining the basic features of the observed $\gamma$-ray light curves, while there seems broad consensus that the particle accelerator is located in the outer magnetosphere, the radiation physics remain controversial.

Recently, motivated by the new discoveries of VHE emission from MAGIC and  especially VERITAS collaborations \cite{2008Sci...322.1221A,VERITASPSRDetection} we argued in favor of inverse Compton origin  of pulsar high energy emission  \cite{2012ApJ...754...33L,2012ApJ...757...88L,2012arXiv1208.5329L}.  Let us here briefly summarize the arguments in favor of IC scattering:

{\bf Maximal energy of curvature emission in Crab}. The curvature emission in pulsars is limited to energies below 
\be
\epsilon_{br} = (3 \pi)^{7/4}  { \hbar \over ( c e) ^{3/4} } \eta^{3/4} \sqrt{\xi} \,  {  B_{NS}^{3/4} R_{NS}^{9/4}\over P^{7/4}} 
\label{1}
\ee
where $R_L$ is the light cylinder radius, $P$ is pulsar period of rotation, $\xi$ is a dimensionless scaling parameter $\xi=R_c/R_L$, $R_c$ is the radius of curvature of \Bf lines, $B = B_{NS} (R_{NS}/R)^3$, where $B_{NS}$ is the magnetic field on the surface of the neutrons star and $R_{NS}$ the starÕs surface  and $\eta = E/B \leq 1$ is the relative  strength of the accelerating \Ef, \citep{2012ApJ...754...33L}. 

\begin{figure}[h!]
\includegraphics[width=0.49\linewidth]{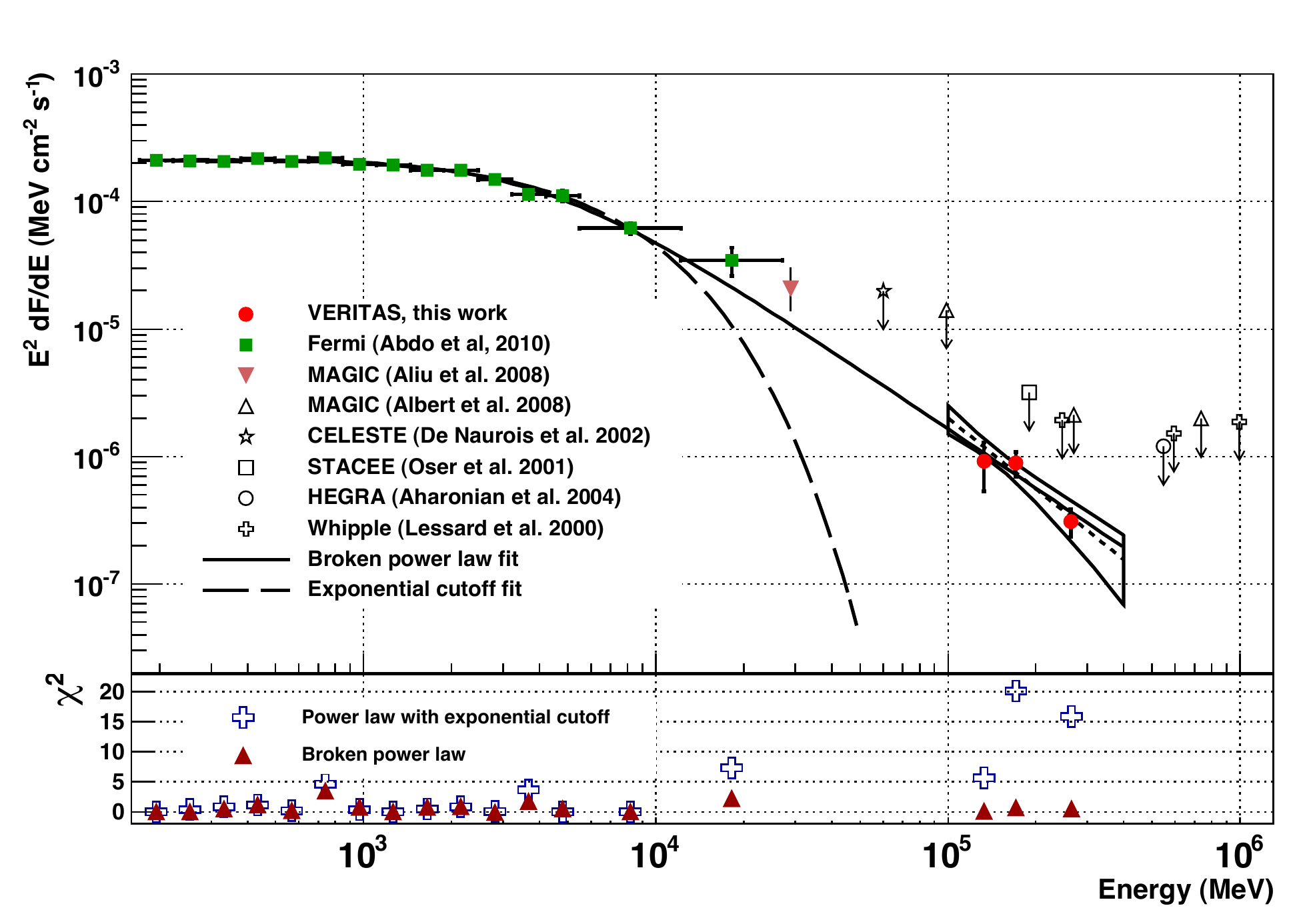}
\includegraphics[width=0.49\linewidth]{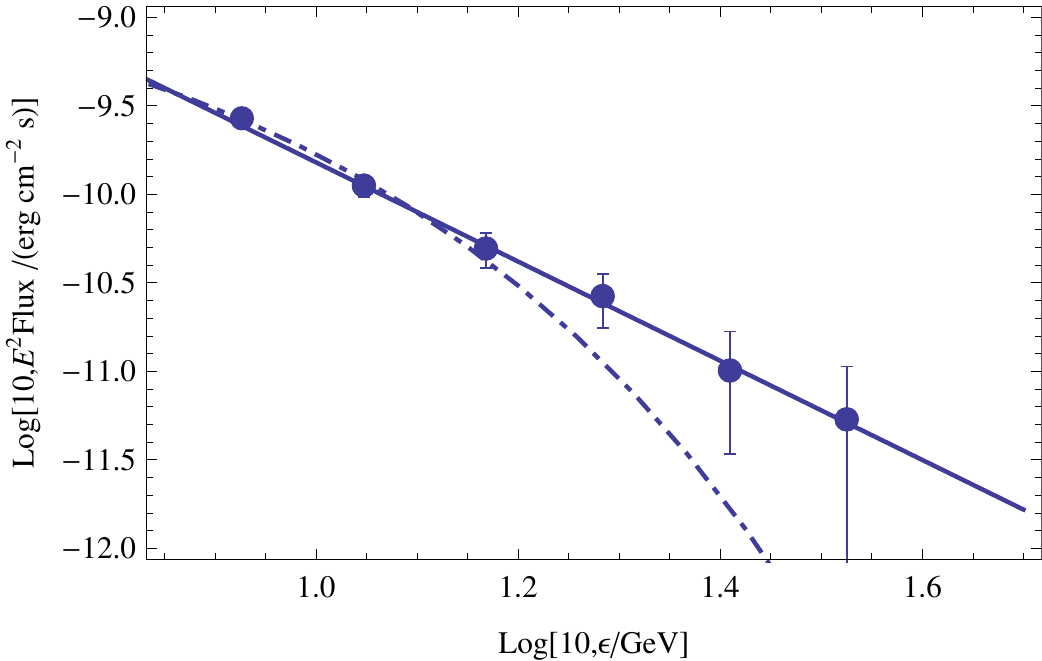}
\caption{Left: High energy spectrum of Crab demonstrating the non-exponential spectral break inconsistent with curvature emission \citep[figure from][]{VERITASPSRDetection} .  Right: Fits to the high energy tail of the  Geminga spectrum: power law (solid line, $\chi^2 =0.1$) and exponential cut-off (dashed line, $\chi^2 =2$). }
\label{crab-spectrum}
\end{figure}

If the $\gamma$-ray photons are due solely to the  curvature emission of a radiation reaction-limited population of leptons, the spectrum above the break must show an  exponential cut-off. 
The detection of the Crab pulsar by VERITAS collaboration   \citep{VERITASPSRDetection} clearly demonstrated the non-exponential cut-off above the spectral break, see Fig. \ref{crab-spectrum}.

\begin{figure}[htb]
\includegraphics[width=0.49\linewidth]{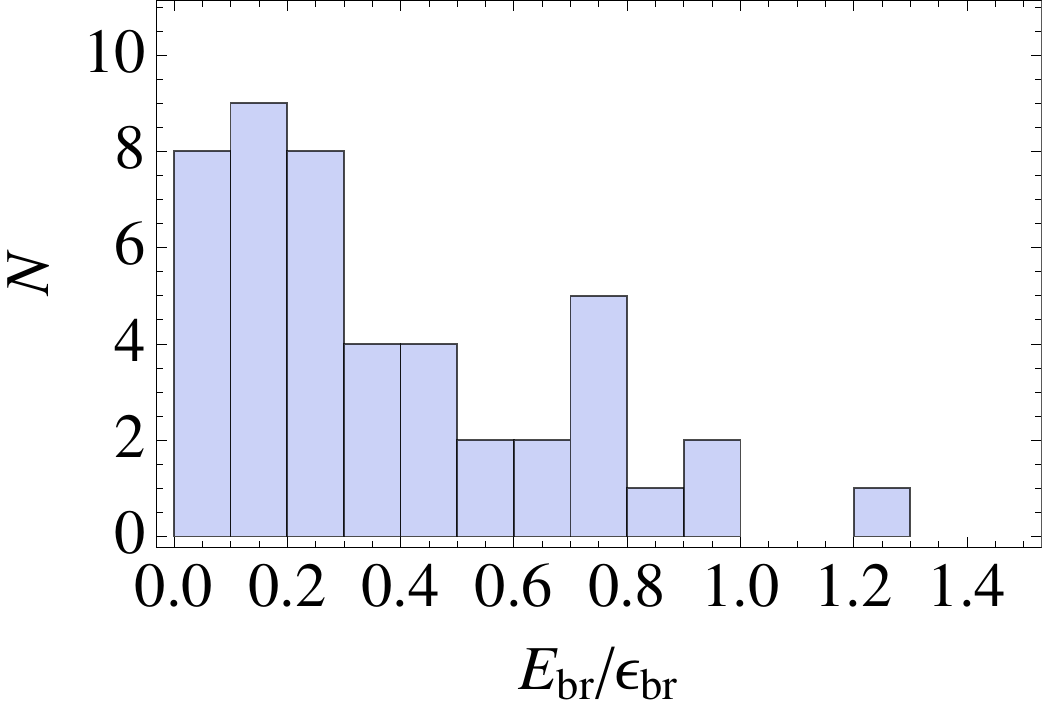}
\caption{Ratio of the  observed break energies $E_{br}$ for 46 pulsars to the maximum predicted for curvature radiation $\epsilon_{br}$, which is given by Eq.\ (\ref{1}) with $\eta =\xi=1$. }
\label{ratio}
\end{figure}

{\bf Maximal energy of curvature emission and observed breaks}.  Lyutikov, Ref. \citep{2012ApJ...754...33L},  compared the observed spectral breaks of Fermi pulsars from the first Fermi catalogue with the predicted breaks due to curvature emission,  Eq.\ (\ref{1}), see Fig. \ref{ratio}.
A significant number of pulsars the ratio is close to one and for one pulsar, PSR  J1836 + 5925, the ratio is even larger than one. In order to explain the spectral break for these pulsars as a result of curvature radiation,  an accelerating \Efs\ should be close to or even larger than the \Bf\ at the light cylinder. What is more, the example of Crab demonstrates  that the spectral break may not be related to the maximal curvature photons.

{\bf  Geminga: non-exponential break}. Reanalyzing  the Fermi
  spectra of the Geminga   pulsar  above the break, Lyutikov, Ref.  \cite{2012ApJ...757...88L}, found that it is well approximated by  a simple power law  without the exponential cut-off, making  Geminga's spectrum  similar   to that of  Crab, Fig. \ref{crab-spectrum}.   Vela's broadband $\gamma$-ray spectrum is equally well fit with both the exponential cut-off and the double power law shapes. 

\begin{figure}[htb]
\includegraphics[width=0.99\linewidth]{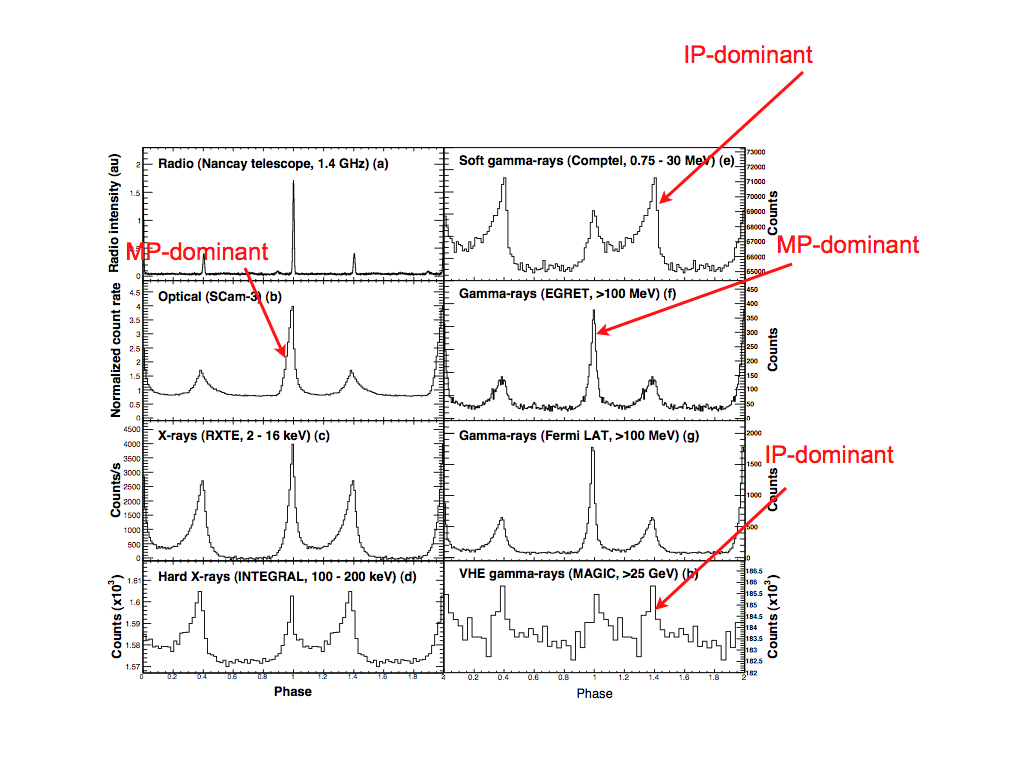}
\caption{Evolution of the Crab profile with energy. Note that that the lower-energy evolution of the increasing interpulse to main pulse ratio is mirrored in the $\gamma$-rays. Such behavior is expected in synchrotron-self-Compton model. }
\label{Crab-profile}
\end{figure}

{\bf Patterns of relative intensities in the Crab  of the leading and trailing pulses repeated in the $X$-ray and $\gamma$-ray regions}, see Fig. \ref{Crab-profile}.

\begin{figure}[h!]
\includegraphics[width=.49\linewidth]{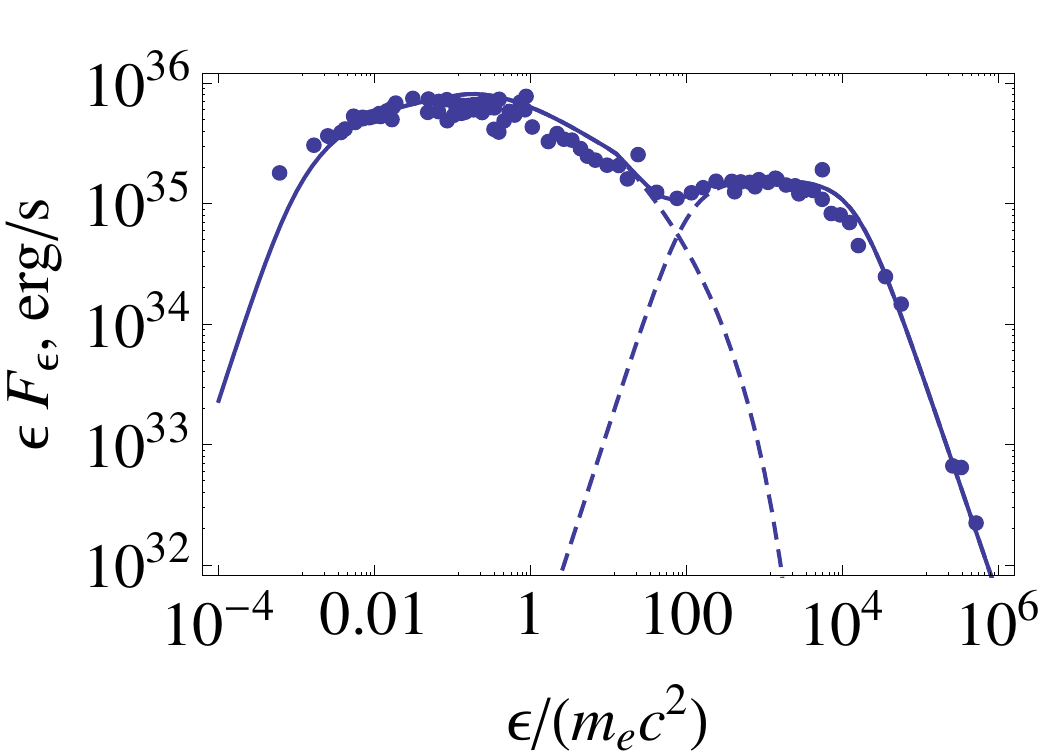}
\includegraphics[width=.49\linewidth]{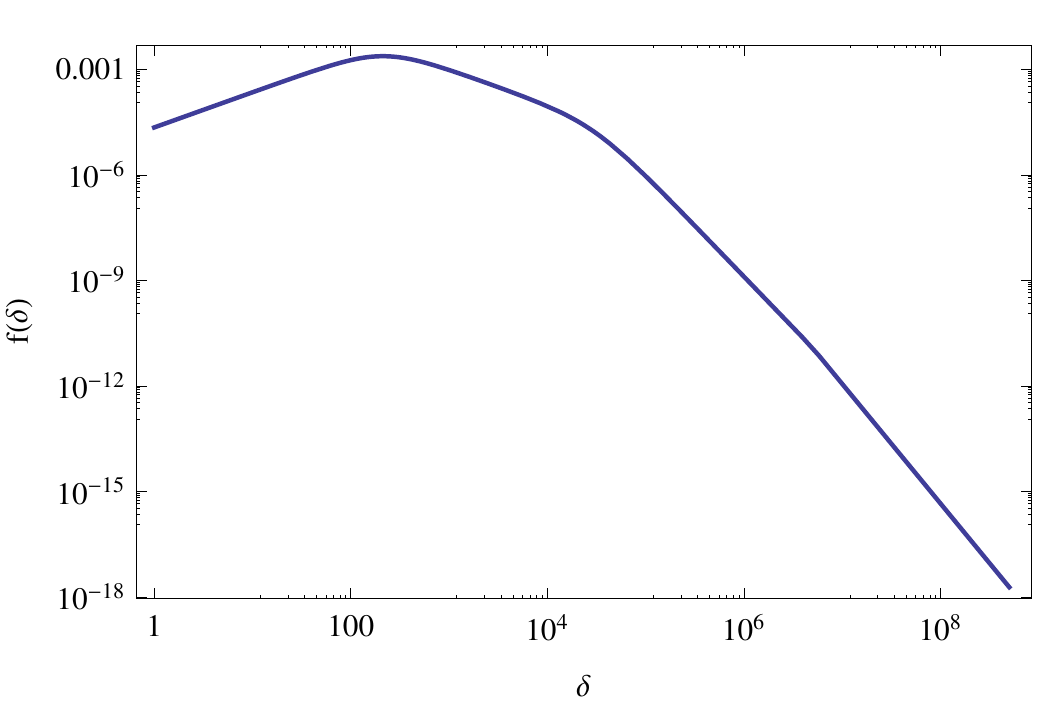}
\caption{Left: The broadband spectrum of the Crab approximated with the CSC model. The IC bump in the KN regime provides a direct measurement of the bulk  particle distribution, while the high energy part of  cyclotron bump constrains the very high energy tail of the particle distribution.
 This is a fit over nearly ten decades in energy, using only a handful of parameters. Right: The  parallel distribution function $f(\delta)$} 
\label{CrabPulsarFit}
\end{figure}

{\bf The broadband  spectrum of  Crab pulsar, from UV to very high energy gamma-rays}  - nearly ten decades in energy - can be reproduced  within the framework of  the cyclotron-self-Compton model. Emission is produced  by  two counter-streaming beams within the outer gaps, at distances above  $\sim$ 20 NS radii. The outward moving beam  produces UV-$X$-ray photons via Doppler-booster cyclotron emission, and GeV photons  by Compton scattering the cyclotron photons produced by the inward going beam. The scattering occurs in the deep Klein-Nishina regime, whereby the  IC component provides a direct measurement of particle distribution within the magnetosphere.   The required plasma multiplicity is high, $\sim 10^6-10^7$, but is consistent with the average particle flux injected into the pulsar wind nebula \cite{2012arXiv1208.5329L}, Fig.  \ref{CrabPulsarFit}. 

These arguments demonstrates that the inverse Compton scattering may be the dominant high energy emission mechanism in majority of pulsars.

\section{Implications}
\begin{itemize}
\item For IC scattering occurring in the Klein-Nishina regime, the particle distribution in the gap does not evolve towards
a stationary distribution and thus is intrinsically time-dependent.
\item
In a   radiation reaction-limited regime of particle acceleration the gamma-ray luminosity $L_\gamma $  scales {\it linearly}  with the pulsar spin-down power $\dot{E}$, $L_\gamma \propto \dot{E}$, and not proportional to $\sqrt{\dot{E}}$ as expected from potential-limited acceleration.
\item  The importance of  Compton scattering in the Klein-Nishina regime also implies the importance of  pair production in the outer gaps. We suggest that outer gaps are important sources  of pairs  in pulsar magnetospheres.
\item Cyclotron motion of particles  in the pulsar magnetosphere may be excited due to  coherent  emission of radio waves by  streaming particles at the anomalous cyclotron resonance, Ref. \cite{1999ApJ...512..804L}. Thus, a whole range of  Crab non-thermal emission, from coherent radio waves to very high energy $\gamma$-rays - nearly eighteen  decades in energy - may be a manifestation of inter-dependent  radiation processes.
\end{itemize}

\bibliographystyle{aipproc}   

\end{document}